\documentclass{sig-alternate-05-2015}
\usepackage{multirow,booktabs}
\usepackage{mathtools}
\usepackage{stmaryrd}
\usepackage{paralist}
\usepackage{url}
\usepackage{relsize}
\usepackage{multirow}
\usepackage{graphicx}
\usepackage{float}
\usepackage{multicol}
\usepackage{subfigure}
\usepackage{hyperref}
\usepackage{makeidx}  
\usepackage{graphicx}
\usepackage{pgfplots}
\usepackage{multirow}
\usepackage{amsmath}
\usepackage{amssymb}
\usepackage{subfigure}
\usepackage{ifsym}
\usepackage{stmaryrd}
\usepackage{algorithm}
\usepackage{algorithmic}
\usepackage{tikz,times}
\usepackage{xcolor}

\usepackage{lipsum}

\usepackage{epstopdf}
\usepackage{epsfig}

\newcommand{\SPARQL}{\mathop{\mathrm{SPARQL}}}

\newcommand{\semm}[2]{\llbracket #1 \rrbracket_{#2}}

\newtheorem{theorem}{Theorem}
\newtheorem{proposition}[theorem]{Proposition}

\usepackage{tikz}
\usetikzlibrary{arrows,positioning,automata,decorations,fit,backgrounds,calc,shapes,snakes}
\usepgflibrary{shapes.geometric} 
\usetikzlibrary{shapes.geometric} 
\usepgflibrary{decorations.pathmorphing} 
\usetikzlibrary{decorations.pathmorphing} 
\usepackage{verbatim}
\usetikzlibrary{calc,backgrounds}
\usetikzlibrary{trees}

\begin{document}

\setcopyright{acmcopyright}

\doi{http://dx.doi.org/10.1145/0000000.0000000}

\isbn{978-1450317412}

\conferenceinfo{AOSD'12}{Hasso-Plattner Institut Potsdam, Germany,March 25--30, 2012}

\acmPrice{\$15.00}

%
\conferenceinfo{XXX}{XXX}
\CopyrightYear{2016} 
\crdata{0-12345-67-8/90/01}  

\title{PRSP: A Plugin-based Framework for RDF Stream Processing}
%
%
%
%
%

\numberofauthors{3} 
%
\author{
%
%
Qiong Li$^{1,3}$ \hspace{0.1cm} Xiaowang Zhang$^{1,3}$  \hspace{0.1cm} Zhiyong Feng$^{2,3}$\\
       \affaddr{$^1$School of Computer Science and Technology,Tianjin University, Tianjin 300350, P. R. China}\\
       \affaddr{$^2$School of Computer Software,Tianjin University, Tianjin 300350, P. R. China}\\
       \affaddr{$^3$Tianjin Key Laboratory of Cognitive Computing and Application, Tianjin 300350, P.R. China}}
%

\maketitle
\begin{abstract}
In this paper, we propose a plugin-based framework for RDF stream processing named PRSP. Within this framework, we can employ SPARQL query engines to process C-SPARQL queries with maintaining the high performance of those engines in a simple way. Taking advantage of PRSP, we can process large-scale RDF streams in a distributed context via distributed SPARQL engines. Besides, we can evaluate the performance and correctness of existing SPARQL query engines in handling RDF streams in a united way, which amends the evaluation of them ranging from static RDF (i.e., RDF graph) to dynamic RDF (i.e., RDF stream). Finally, within PRSP, we experimently evaluate the correctness and the performance on YABench. The experiments show that PRSP can still maintain the high performance of those engines in RDF stream processing although there are some slight differences among them.
\end{abstract}

\printccsdesc

\terms{Theory}
\vspace*{-8pt}
\keywords{RDF Stream; RSP; SPARQL; C-SPARQL}

\vspace*{-8pt}
\section{Introduction}\label{sec:intro}

RDF stream, as a new type of dataset, can model real-time and continuous information in a wide range of applications, e.g. environmental monitoring, Smart City and so on. But data stream is unbounded sequences of time-varying data element and difficult to store.

What is more, there is a few RSP\cite{rsp} (RDF Stream Processing) systems, such as C-SPARQL\cite{csparql} and EP-SPARQL\cite{epsparql} implemented for supporting RDF stream due to its complicacy in processing. On the other hand, there are many popular and efficient SRARQL query engines supporting only static RDF graphs, such as the centralized engines, Jena\cite{Jena}, RDF-3X and gStore, and distributed systems, TriAD, gStoreD\cite{gStoreD} and so on. How to employ those SRARQL query engines to evaluate continuous queries becomes an interesting problem.

In this paper, we provide a plugin-based framework for RDF stream processing named PRSP, which makes it possible to use the high-performance RDF engines that's valid only for RDF graphs, to process RDF streams. Moreover, within this framework, we can employ any RDF query engine to process RDF streams in a convenient way and compare their performance under a unified framework namely PRSP. And users can choose the favourable systems based on their all kinds of requirements. For example, they have the need to handle large-scale RDF graphs, thus distributed engines are the best choice.

\vspace*{-8pt}
\section{Preliminaries}\label{sec:DataQuery}

\textbf{RDF stream} An RDF stream is defined as ordered sequences of pairs, each pair being made of an RDF triple and a timestamp $T$:
$$(\langle S_{i}, P_{i}, O_{i}\rangle , T)$$

\textbf{C-SPARQL query} The continuous query is divided into three parts: $R_{\text{\text{query}}}$, $S(t)$, $Q_{\SPARQL}$, and it is formally defined as follows:
\hspace*{0.75 in}   $Q = [R_{\text{\text{query}}}, S(t), Q_{\SPARQL}]$

\begin{compactitem}
    \item $R_{\text{\text{query}}}$ indicates the registered query from users which is waiting to be addressed.
	\item $S(t)$ is the RDF stream registered by the RSP systems, which defines the window size and step size.
    \item $Q_{\SPARQL}$ is a standard RDF query language, i.e., SPARQL.
\end{compactitem}

\vspace*{-6pt}
\begin{proposition}\label{prop:sound}
Let $Q$ be a C-SPARQL query. For any RDF stream $S$ and any present time $t$, the following holds:
$$\semm {Q}{(S, t)} = \semm {Q_{\SPARQL}}{\text{Window}(S, t)}.$$
\end{proposition}
\vspace*{-6pt}

Proposition \ref{prop:sound} ensures that  the evaluation problem of C-SPARQL queries over RDF streams can be equivalent to the evaluation problem of SPARQL queries over RDF graphs. Moreover, Proposition \ref{prop:sound} can show that the evaluation problem of C-SPARQL has the same computational complexity as  SPARQL \cite{csparql}.

Consider the following query, a simple example from C-SPARQL. Line 1 matching the $R_{\text{query}}$, tells the RSP system to register the continuous query of \emph{TestQuery}. $S(t)$, that is the following list of line 3, indicates that \emph{streams} with a sliding window of 5 seconds that slides every 5 seconds, is the stream data waiting to be processed. And $Q_{sparql}$, displayed in both line 2 and line 4, is the query language for RDF.

\vspace*{-4pt}
\begin{center}
\fbox{\shortstack[l]{
1. \textbf{REGISTER QUERY} \emph{TestQuery} AS \\
2. \textbf{SELECT} ?obs \\
3. \textbf{FROM STREAM} \emph{streams} [ \textbf{RANGE} 5s \hspace{0.1cm} \textbf{STEP} 5s ]\\
4. \textbf{WHERE} \{ ?obs observedProperty AirTemperature. \}
}}
\end{center}

\vspace*{-8pt}
\section{The architecture of PRSP}\label{sec:architecture}
\begin{figure}[h]
\centering
\includegraphics[width=3.4in]{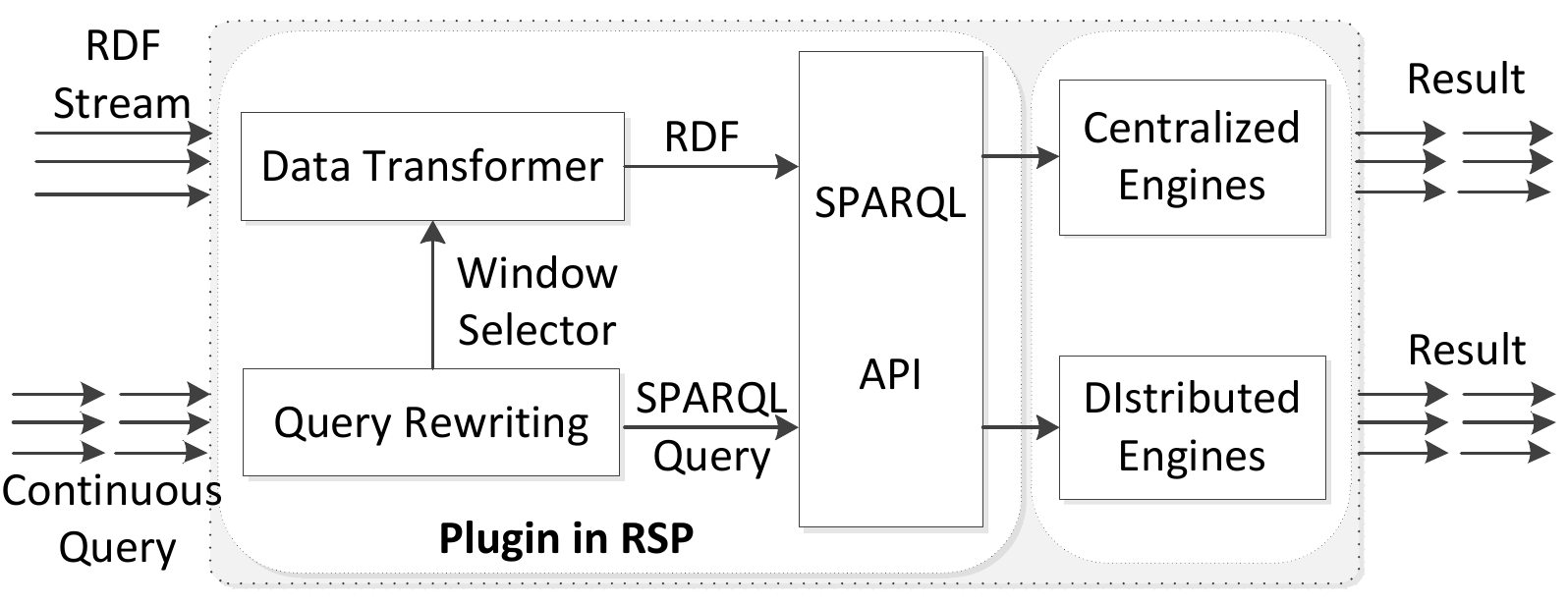}\\
\vspace*{-8pt}
\caption{PRSP Architecture}\label{fig:architecture}
\vspace*{-15pt}
\end{figure}

PRSP is an extension of SPARQL for querying both RDF graphs and RDF streams shown in Figure\ref{fig:architecture}. Both continuous query and RDF streams as the input of PRSP are transformed by the plugin Query Rewriting and Data Transformer in PRSP, respectively. After that, the output from the former plugins as the input of SPARQL API, the results are produced by one of SPARQL engines. 
And the right box consists of any SPARQL query engine which is used as a black box for evaluating RDF graphs. Its architecture contains three types of plugin: Data Transformer, Query Rewriting, and a SPARQL API connecting with the former two plugins.


\vspace*{-1pt}
\paragraph{Query Rewriting} Continuous queries as the input of query rewriting mode, apply transformation methods in order to generate two types of queries, namely, SPARQL query and window operator, which can be addressed in one of SPARQL engine and Data Transformer module, respectively. After rewritting $Q$, we can obtain $Q_{\SPARQL}$.

\vspace*{-8pt}
\paragraph{Data Transformer} The data transformer module manages RDF streams specified in the query via Esper or another DSMS. And it transforms RDF streams into RDF graphs based on the window size and step size set by window operator. After tranforming $S$ w.r.t. $t$, we can obtain $\text{Windows}(S, t)$.

\vspace*{-8pt}
\paragraph{SPARQL API} PRSP defines a unified interface for RDF engines, which makes it possible and easy for SPARQL engines to process RDF streams. In the current version of PRSP, we have extended PRSP by including a few centralized engines, such as Jena, gStore, and RDF-3X, and two distributed engines, i.e., gStoreD and TriAD.

\vspace*{-3pt}
\section{Experiments and Evaluations}\label{sec:experiments}

\vspace*{-10pt}
\paragraph{Experiments} All centralized experiments were carried out on a machine running Linux, which has 4 CPUs with 6 cores and 64GB memory, and 5 machines with the same performance for distributed experiments. For evaluation, we utilized YABench RSP benchmark\cite{yabench}, which uses a real world dataset about water temperature. In our experiments, we performed sliding windows with a window size and a step size of 5 seconds, respectively. Considering that some engines can not support complex queries, the experiments used two BGP queries, $Q1$ and $Q1'$. $Q1$ is a BGP query with four forms($Q1$) from YABench, and $Q1'$ is the rewriting of $Q1$ with three triples. Since RDF-3X did not work when the the amount of stream data to 42.000 triples (i.e.,$s=500$), we chose five load scenario (i.e., $s =100/200/300/400/500$ $sensors$).

\vspace*{-8pt}
\begin{figure}[h]
\subfigure[The response time of Q1]{
\begin{minipage}[t]{0.48\linewidth}
\centering
\scalebox{0.5}{
\begin{tikzpicture}
\begin{semilogyaxis}[
    xlabel={The number of sensors},
    ylabel={Time[ms]},
ymin=0,
    symbolic x coords={100,200,300,400,500},
    ymajorgrids=true,
    grid style=dashed,
    anchor=north,
    legend pos = north west,
    ymax=10000,
	xlabel style={below=-0.02cm},
    legend style={at={(-0.125, -0.2)}}, 
    legend columns=-1, 
    ylabel near ticks,
]

\addplot[
    color=blue,
    mark=square*, mark options={fill=black}
    ]
    coordinates {
    (100,196)(200,236)(300,245)(400,314)(500,269)
    };

\addplot[
    color=black,
    mark=triangle*, mark options={fill=white}
    ]
    coordinates {
    (100,12)(200,13)(300,23)(400,27)(500,10000)
    };

\addplot[
    color=blue,
    mark=square*, mark options={fill=red}
    ]
    coordinates {
    (100,175)(200,435)(300,876)(400,843)(500,1117)
    };

\addplot[
    color=orange,
    mark=*, mark options={fill=blue}
    ]
    coordinates {
    (100,4393)(200,4387)(300,4369)(400,4445)(500,4339)
    };

 \addplot[
    color=orange,
    mark=*, mark options={fill=green}
    ]
    coordinates {
    (100,90)(200,118)(300,180)(400,186)(500,221)
    };
    \legend{Jena,RDF3X,gStore,gStoreD,TriAD}

\end{semilogyaxis}
\end{tikzpicture}
}
\end{minipage}
}
\subfigure[The response time of Q1']{

\begin{minipage}[t]{0.48\linewidth}
\centering
\scalebox{0.5}{
\begin{tikzpicture}
\begin{semilogyaxis}[
    xlabel={The number of sensors},
    ylabel={Time[ms]},
    ymin=0,
    symbolic x coords={100,200,300,400,500},
    ymajorgrids=true,
    grid style=dashed,
    anchor=north,
    legend pos = north west,
    ymax=10000,
	xlabel style={below=-0.02cm},
    legend style={at={(-0.126, -0.2)}}, 
    legend columns=-1, 
ylabel near ticks,
]

\addplot[
    color=blue,
    mark=square*, mark options={fill=black}
    ]
    coordinates {
    (100,178)(200,224)(300,242)(400,240)(500,283)
    };

\addplot[
    color=black,
    mark=triangle*, mark options={fill=white}
    ]
    coordinates {
    (100,11)(200,19)(300,26)(400,25)(500,83)
    };

\addplot[
    color=blue,
    mark=square*, mark options={fill=red}
    ]
    coordinates {
    (100,294)(200,501)(300,633)(400,689)(500,801)
    };

\addplot[
    color=orange,
    mark=*, mark options={fill=blue}
    ]
    coordinates {
    (100,4993)(200,4387)(300,4369)(400,4445)(500,4339)
    };

 \addplot[
    color=orange,
    mark=*, mark options={fill=green}
    ]
    coordinates {
    (100,147)(200,171)(300,187)(400,202)(500,213)
    };
    \legend{Jena,RDF3X,gStore,gStoreD,TriAD}

\end{semilogyaxis}
\end{tikzpicture}
}
\end{minipage}
}
\vspace*{-8pt}
\caption{Querying time in different scenarios within PRSP}\label{fig:queryTime}
\end{figure}
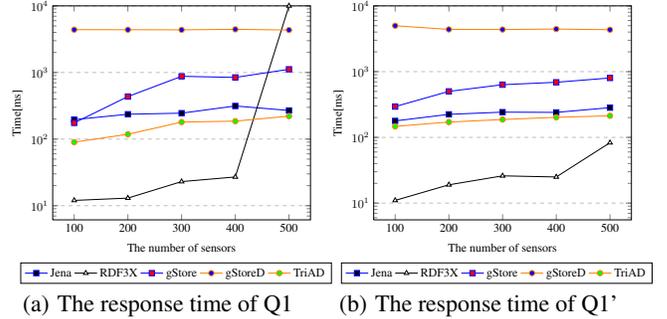

\begin{figure}[htbp]
\subfigure[s = 100 sensors]{
\begin{minipage}[t]{0.48\linewidth}
\centering

\scalebox{0.51}{
\begin{tikzpicture}

\begin{semilogyaxis}[
	bar width=6pt,
	symbolic x coords={LT,RT,ET},
	ylabel=Time(ms),
	enlarge x limits=0.45,
	ymin=0,ymax=20000,
	legend style={at={(1, -0.15)}, 
    legend columns=-1, 
    ylabel near ticks,
    },
	ybar,
]
\addplot
	coordinates {(LT,55) (RT,178) (ET,252)};
\addplot
	coordinates {(LT,557) (RT,11) (ET,868)};
\addplot
	coordinates {(LT,5783) (RT,294) (ET,6049)};
\addplot
	coordinates {(LT,10313) (RT,4393) (ET,15095)};
\addplot
	coordinates {(LT,32) (RT,147) (ET,9290)};
\legend{Jena, RDF3X, gStore, gStoreD, TriAD}
\end{semilogyaxis}
\end{tikzpicture}
}
\end{minipage}
}
\subfigure[s = 500 sensors]{
\begin{minipage}[t]{0.44\linewidth}
\centering
\scalebox{0.51}{
\begin{tikzpicture}
\begin{semilogyaxis}[
	bar width=6pt,
	symbolic x coords={LT,RT,ET},
	ylabel=Time(ms),
	enlarge x limits=0.45,
	ymin=0,ymax=20000,
	legend style={at={(1, -0.15)}, 
    legend columns=-1, 
    ylabel near ticks,
    },
	ybar,
]
\addplot
	coordinates {(LT,104) (RT,283) (ET,423)};
\addplot
	coordinates {(LT,862) (RT,83) (ET,1705)};
\addplot
	coordinates {(LT,6736) (RT,909) (ET,7672)};
\addplot
	coordinates {(LT,11044) (RT,4339) (ET,16251)};
\addplot
	coordinates {(LT,31) (RT,213) (ET,9787)};
\legend{Jena, RDF3X, gStore, gStoreD, TriAD}

\end{semilogyaxis}
\end{tikzpicture}
}
\end{minipage}
}
\vspace*{-8pt}
\caption{RDF stream for processing time in PRSP}\label{fig:LT_RT_ET}

\end{figure}
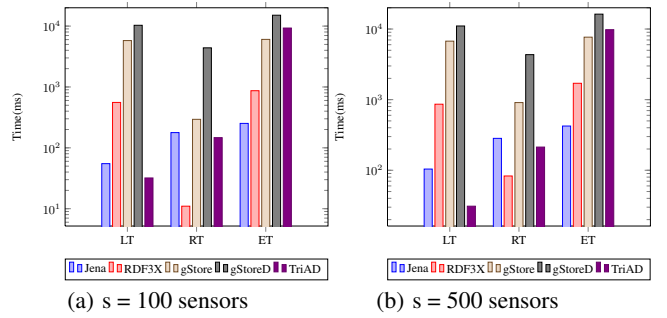

\vspace*{-3pt}
\paragraph{Evaluations} The performance of each engine under the five different input loads for windows is shown in Fig. \ref{fig:queryTime}. When the load ranges from $s=100$ to $s=500$, the query response time is with varying degrees of increase except for gStoreD. Fig. \ref{fig:LT_RT_ET} shows the time of three processes, including data load time ($LT$), query response time($RT$), and engine execution time ($ET$) under $s=300$ obtained from query $Q1'$. $LT$ from RDF3X, gStore, and gStoreD occupies a large part of $ET$, resulting in their lower efficiency for processing RDF streams. Table \ref{tab:P/C} illustrates the results of precision and recall from the experiments under three load scenarios (i.e., $s=100/300/500$) in PRSP. Along with more input load for windows, most of them enjoy lower recalls with high accuracy.

\vspace*{-10pt}
\vspace*{-20pt}

\begin{table}[H]
\centering
\vspace*{-11pt}
\caption{Precision/Recall results }\label{tab:P/C}
\vspace*{4pt}
\begin{tabular} {|c|c|c|c|c|c|c|}
\hline

 &  & Jena & RDF3X & gStore & gStoreD & TriAD\\ \hline
\multirow{3}{*}{Precision} &s=100 &99\% &93\% &100\%  &100\% &97\%\\ \cline{2-7}
                           &s=300 &97\% &94\% &100\% &100\% &93\%\\ \cline{2-7}
                           &s=500 &85\% &88\%    &100\% &100\% &100\%\\ \hline

\multirow{3}{*}{Recall}    &s=100 &95\% &89\% &75\% &72\% &95\%\\ \cline{2-7}
                           &s=300 &94\% &91\% &88\% &76\% &92\%\\ \cline{2-7}
                           &s=500 &92\% &79\%  &77\%  &63\%  &91\%\\ \hline
\end{tabular}
\vspace*{-8pt}
\end{table}

\vspace*{-10pt}
\section{Conclusions}\label{sec:conclusion}
In this paper, we present PRSP, as a plugin adaptable for SPARQL engines, to process RDF streams, which makes it feasible to employ various engines to process large-scale RDF streams. In the future, we will optimize PRSP further to improve its performance and correctness. This work is supported by the National Key Research and Development Program of China (2016YFB1000603) and the National Natural Science Foundation of China (61672377).

%

\vspace*{-8pt}
\begin{thebibliography}{10}

\bibitem{rsp}
\newblock {\em http://www.w3.org/community/rsp/.}

\bibitem{csparql}
Barbieri, D. F., Braga, D., Ceri, S., Della Valle, E., and Grossniklaus, M.
\newblock {\em Querying RDF streams with C-SPARQL}.
\newblock ACM SIGMOD Record, 2010, 39(1): 20-26.

\bibitem{epsparql}
Anicic D, Fodor P, Rudolph S, et al.
\newblock {\em EP-SPARQL: a unified language for event processing and stream reasoning.}
\newblock In: Proc. of WWW'11, pp. 635--644.

\bibitem{Jena}
\newblock {\em http://jena.sourceforge.net/ARQ.}

\bibitem{gStoreD}
P.~Peng, L.~Zou, MT.~Zsu, L.~Chen, and D.~Zhou.
\newblock  {\em Processing SPARQL queries over distributed RDF graphs.}
\newblock  VLDB J., 2016, 25(2): 243--268.


\bibitem{yabench}
Kolchin M, Wetz P, Kiesling E, et al.
\newblock {\em YABench: A Comprehensive Framework for RDF Stream Processor Correctness and Performance Assessment}.
\newblock In: Proc. of ICWE'16, pp.280-298.




\end{thebibliography}
%
%

\end{document}